\documentclass[superscriptaddress,preprint]{revtex4}
\usepackage{graphicx}
\usepackage{epstopdf}
\usepackage{amsmath}
\usepackage{color}
\usepackage{enumitem}
\usepackage{color}
\usepackage{verbatim}
\begin{document}
%\draft
\title{How soft repulsion enhances the depletion mechanism}
\author{Lorenzo Rovigatti} 
\affiliation{Faculty of Physics, University of Vienna, Boltzmanngasse 5, A-1090 Vienna, Austria}
\affiliation{Dipartimento di Fisica, {\em Sapienza} Universit\`a di Roma, Piazzale A. Moro 2, 00185 Roma, Italy}
\author{Nicoletta Gnan}
\affiliation{CNR-ISC, Uos Sapienza, Piazzale A. Moro 2, 00185 Roma, Italy}
\author{Alberto Parola}
\affiliation{Dipartimento di Scienza e Alta Tecnologia, Universit\`{a} dell'Insubria, Via Valleggio 11, 22100 Como, Italy}
\author{Emanuela Zaccarelli} 
\affiliation{CNR-ISC, Uos Sapienza, Piazzale A. Moro 2, 00185 Roma, Italy}
\affiliation{Dipartimento di Fisica, {\em Sapienza} Universit\`a di Roma, Piazzale A. Moro 2, 00185 Roma, Italy}
\begin{abstract}
We investigate binary mixtures of large colloids interacting through soft potentials with small, ideal depletants.
We show that softness has a dramatic effect on the resulting colloid-colloid effective potential when the depletant-to-colloid size ratio $q$ is small, with significant consequences on the colloidal phase behaviour. We also provide an exact relation that allows us to obtain the effective pair potential for {\it any} type of colloid-depletant interactions in the case of ideal depletants, without having to rely on complicated and expensive full-mixture simulations. We also show that soft repulsion among  depletants further enhances the tendency of colloids to  aggregate.
Our theoretical and numerical results demonstrate that --- in the limit of small $q$ --- soft mixtures cannot be mapped onto hard systems and hence soft depletion is not a mere extension of the widely used Asakura-Oosawa potential. \end{abstract}

\pacs{}

%\date{\today}

\maketitle

Colloidal particles can be considered as super-atoms~\cite{poon_big_atoms}, moving in a background fluid, and are usually described in terms of effective interactions~\cite{Likos}. The latter are not fixed by chemistry but can be tuned almost arbitrarily by a careful design of the suspension~\cite{Glotz_Solomon_natmat}, making soft matter systems display states and phases that have no counterpart in atomic and molecular systems~\cite{And02a,granicknature,lenz_prl,closed_loop_prl,biffi_dna,dna_tetra_gel}.
As an example, the polymeric nature~\cite{likos_star_polymers,barbara_telechelic,rings} of some colloidal systems allows for a fine tuning of their molecular architecture as well as their softness.
Indeed, while hard-sphere (HS) colloids such as PMMA particles have become a favourite model system to study phase transitions and dynamics~\cite{pusey_nature_1986}, more recently soft colloids have gained increasing attention. Among these, microgel PNIPAM particles have emerged as a prototype for soft repulsive colloids~\cite{paloli_sm,ballauff_soft_matter,han_microgel}. 
While a consensus has not been reached yet about microgel effective interactions, comparisons with experiments have shown that they can be described by soft potentials~\cite{heyes_microgel,richtering_microgel,mohanty_pnipam,riest2012coarse}.

In addition to the possibility of changing the nature of colloidal particles themselves, a well-established way to tune colloidal interactions is to add a co-solute to the suspension, often in the form of non-adsorbing polymers~\cite{Poo93b} or surfactants~\cite{buzzaccaro_B2}. The resulting \textit{depletion} forces, controlled by the size and the concentration of the additives (also called depletants), give rise to an effective colloid-colloid attraction.
Depletion interactions have been known for about sixty years since the pioneering works of 
Asakura and Oosawa (AO)~\cite{original_AO} and Vrij~\cite{vrij_AO}. In the case of  a mixture of HS colloids and ideal polymers, with polymer-colloid HS interactions, the effective colloid-colloid potential can be derived analytically. This AO formulation has become the reference model system for depletion interactions~\cite{Likos} and its use is widespread~\cite{mao1995depletion,dijkstra_1999,LuCiulla,Buz10,NatGnan,binder_ao_review}.
For small depletant-to-colloid size ratio $q$ 
the most unusual features of colloidal behaviour arise. Among these, we mention the metastability of the gas-liquid spinodal~\cite{And02a}, which enhances nucleation~\cite{tenwolde} and gelation~\cite{LuCiulla}, and the appearance of two different glasses separated by a reentrant liquid at packing fractions larger than a simple HS glass~\cite{pham_multiple_glassy_states}. On top of this, depletion plays an important role also in biological systems, from the folding of single 
biopolymers~\cite{minton_biopolymers,snir_helix_science,hansen-goos_prl} to systems where macromolecular crowding 
can significantly impact both structure and kinetics~\cite{minton_biopolymers,ellis_bio_depletion,sarkar_bio_depletion,sapir}.

In the last decade, considerable efforts have been devoted to the study of non-ideal colloid-polymer mixtures, building on the 
simple AO model with the aim of improving the description of polymer-polymer and polymer-colloid interactions.
Obtaining reliable effective potentials under these conditions requires sophisticated theoretical and numerical calculations~\cite{bolhuis_prl,bolhuis_macromol,bolhuis_jcp,cinacchi2007,zausch2009,wilding_2011,majka_2013}.
%In addition, the study of mixtures of star polymers and linear chains has predicted the occurrence of anomalous depletion effects~\cite{camargo_likos}. 
Recent work has also been devoted to the effects induced by interactions between depletant molecules, such as the investigation of the phase behaviour of non-ideal mixtures of hard spheres with very short-range Yukawa tails~\cite{mixtures_hs} and a few studies on the effect of attractive interactions on depletion forces~\cite{amokrane_1,amokrane_2,grest}. Moreover, when colloid-depletants interactions are not hard, as in all these systems, it has been shown that the depletion 
mechanism, which in the AO model is controlled solely by entropy, can become dominated by enthalpy~\cite{harries_review,jeetain_crowding,sapir}.

Even though a qualitative difference between the AO model and depletion effects in non-hard systems has already 
been noted~\cite{mixtures_hs,cinacchi2007,camargo_likos,sapir}, the generalisation of the AO mixture to the case of soft colloid-depletant 
interactions has not been thoroughly tested, except for the work of Zausch and coworkers~\cite{zausch2009}, who have shown that 
for $q=0.8$ the system can be assimilated to a HS mixture with an effective particle diameter.
In this work we take a step forward and investigate the effects of softness in colloid-depletant interactions.
Treating the depletant molecules as ideal,  we show that even a small amount of softness produces a dramatic effect on the resulting depletion attraction at small $q$-values. By focusing on representative model systems for soft potentials, such as inverse power-law and hertzian potential, we calculate numerically and theoretically the effective interaction between soft colloids 
immersed in ideal depletants, in the case of colloid-depletant soft repulsion. 
We find that the resulting depletion attraction is strongly enhanced with respect to the AO case both in range and in depth, a feature that is generic for any kind of soft interaction. By comparing soft depletion with the corresponding AO case in terms of second virial coefficients, we show that, unlike one-component systems~\cite{wca_mapping,inverse_B2}, soft mixtures cannot be mapped onto hard ones. These results hold true also when interactions between depletant particles in the form of soft repulsion are considered.
Our results have, thus, profound consequences on the phase behaviour of depletion-interacting soft colloids, which 
have not been really appreciated so far, except for few sporadic studies~\cite{cinacchi2007,majka_2013}.
 
\section{Methods}

\subsection{Theory}
Without loss of generality, the solvent-mediated effective pair potential can be formally expressed in 
terms of grand canonical averages in a pure depletant reservoir at fixed temperature $T$, volume $\mathcal{V}$ and
activity $z_d$~\cite{evans_1999}.
In the special case of {\it ideal} depletants (but arbitrary colloid-depletant interaction), this
expression can be written in closed form as
\begin{equation}
\beta\,V_{\rm depl}(R)=\rho_d\int\,d{\bf r} \left [ 1-e^{-\beta\,v_{cd}(r)}\right ] 
\left [ e^{-\beta\,v_{cd}(|{\bf R}-{\bf r}|)}-1\right ]
\label{eq:veff}
\end{equation}
where $\rho_d$ is the reservoir depletant number density, $v_{cd}(r)$ is the colloid-depletant potential, $\beta = 1 / k_BT$ and $k_B$ is the Boltzmann constant. Equation~\ref{eq:veff} reduces to the well-known AO potential between two colloids whose centres are at distance $R$ 
when $v_{cd}$ is a pure hard-core interaction. By use of the convolution theorem in Eq. (\ref{eq:veff}),  $V_{\rm depl}$ can be easily evaluated numerically by Fourier transform and added to the direct colloid-colloid interaction $V_{cc}$, to yield the total colloid-colloid potential $V_{\rm tot}=V_{cc}+V_{\rm depl}$.

As representative models of soft spheres, we focus on interactions (both for colloid-colloid and for colloid-depletant) given by (i) an inverse power-law potential with exponent $n$, $V_{n}(r) =\epsilon \left( \sigma_{\mu}/ r\right)^n$, where $\epsilon=1$ is the energy strength in units of $k_BT$ 
and (ii) a hertzian potential, $V_{HZ} = \lambda \left(1 - r/\sigma_{\mu} \right)^{5/2}\Theta(\sigma-r)$, where $\Theta$ is the Heaviside step function and $\lambda=500k_BT$ is the strength of the interaction (fixed by recent comparison with experiments\cite{paloli_sm}). Here the subscript $\mu$ refers to the three length-scales of the problem $\sigma_c$, $\sigma_d$ and $\sigma_{cd}$, where  $\sigma_{cd} = \frac{\sigma_c+\sigma_d}{2}$.  A key role is played by the depletant-to-colloid size ratio, formally defined as $q=\sigma_d/\sigma_c$. 
As shown in the Appendix, for a steep power-law colloid-depletant interaction ($n\gg 1$),  the resulting effective interaction between two colloids can be approximated as
\begin{eqnarray}
\beta V_{{\rm depl},n}(R)\approx -\frac{2\pi\rho_d}{R}\,
\begin{cases}
Q_1 & {\rm for} \quad R < 2\sigma_{cd}
\cr
Q_2\,e^{-\alpha\,R} & {\rm for} \quad R > 2\sigma_{cd}
\end{cases}
\label{eq:analytic}
\end{eqnarray}
where $\alpha= \frac{n}{\sigma_{cd}}$ and $Q_i$ are fourth degree polynomials in $R-2\sigma_{cd}$. 
This analytical result shows that an approximately exponential tail, absent in the celebrated AO expression, is generated by the softness of the direct interaction.

\subsection{Soft-to-hard mapping}
In order to compare $V_{\rm depl}$ with the AO expression for hard colloid-depletant interactions, 
it is convenient to first map the colloid-colloid interaction into an effective hard-sphere potential. As usual, we define an effective colloidal hard-core diameter $\sigma^c_{\rm eff}=h \sigma_c$ by 
imposing the equality of second virial coefficients~\citep{wca_mapping,inverse_B2}. Such an equivalence is known to faithfully reproduce the properties of the pure colloidal particle suspension at low density. We now ask whether an analogous mapping can be carried out for the colloid-depletant interaction, thereby reducing the system to the well studied AO model.

The second virial coefficient 
\[B_2(V)=-2\pi \int_0^{\infty} [\exp(-\beta V(r)) -1] r^2 dr\]
allows to compare different potentials $V$ by means of a single parameter, which provides a measure of the two-body potential strength.
It normally depends on temperature or, in the case of depletion interactions, on depletant density.
It is now well established, thanks to the work of Noro and Frenkel~\cite{noro_frenkel}, that
the thermodynamic and static properties of 
a wide class of different potentials, 
including a hard-core repulsion plus a short-range attraction, 
are identical when different systems are compared at the same normalized second virial coefficient $B^*_2(V)=B_2/B_2^{HS}(\sigma)$, where $B_2^{HS}(\sigma)=2\pi\sigma^3/3$ and $\sigma$ the HS diameter.
Thus, we map the soft repulsion onto the HS system through the definition of 
an effective hard sphere diameter $\sigma_{\rm eff}$ such that
$$
B_2(V)=B_2^{HS}(\sigma_{\rm eff}).
$$
The use of this formula yields $\sigma_{\rm eff}=h \sigma$, where $h=1.01818$ for $V_{36}$ and
$h=0.9272$ for $V_{HZ}$ with $\lambda=500k_B T$.
In order to avoid the introduction of too many length scales, we adopt throughout the manuscript the convention that,  %as stated before,  
$\sigma_c, \sigma_d, \sigma_{cd}$ identify the characteristic lengths defining the soft particles. These are indeed the quantities directly accessible experimentally, e.g. for hertzian particles they correspond 
to the experimentally determined diameters (by means, for example, of dynamic light scattering~\cite{mohanty_pnipam}). 
When we compare to the AO case we then use the corresponding rescaled effective diameters, multiplying them by the factor $h$, which is 
potential-dependent. In the case of hard particles (and thus for the AO case) $h=1$, by definition. Note that mapping $V_{cc}$ onto an effective hard-core potential requires a rescaling of $\sigma_c$, while to map the depletant density or the AO range we need to rescale $\sigma_d$. 

\subsection{Simulations}

To evaluate effective potentials we perform parallel runs of Monte Carlo (MC) simulations in the canonical ensemble of two large colloids of diameter $\sigma_c$ in solution with small particles of size $\sigma_d$. We use umbrella sampling to constrain the distance between the two colloids~\cite{umbrella_sampling}: in each run, the two large particles can explore only a limited range $\Delta_i$ of reciprocal distances, and the probability $P(x,\Delta_i)$ to find the colloids at a given surface-to-surface distance $x$ within such interval is computed. Since different runs are allowed to have a small overlap in the probed $\Delta_i$, we obtain the total $P(x)$ by merging together all the $P(x,\Delta_i)$ by means of a least-squares-based algorithm. Finally, we extract the effective potential from the relation $\beta V_{\rm depl} = -\ln(P(x))+C$, where $C$ is a constant that is set by imposing $V_{\rm depl} (\infty) = 0$. We explore different cases in which the colloids interact between them and with the co-solute particles with different interaction potentials. 

To test the validity of the calculated effective two-body potentials, we also perform MC simulations of a monodisperse system of colloids interacting with $V_{\rm tot}$ (thereby neglecting many body interactions) and compare them with the corresponding full binary mixture. Due to the large number of depletant particles, especially at the small $q$-values studied here, we use brownian dynamics on GPUs to simulate the full mixture.
 
For the one-component system we simulate $N_c=10000$ colloids  at a density
$\rho\sigma_{c}^3h^3=0.1$, interacting through $V_{\rm tot} = V_{36} + V_{\rm depl}$, with
$V_{\rm depl}$ obtained from Eq.~\ref{eq:veff}. For comparison, we also simulate the same system interacting with the AO potential given by
\begin{equation}
\beta V_{AO}(r) = -\rho_d \frac{4\pi}{3} \sigma_{cd}^3h^3 \left[ 1 - \frac{3}{4} \frac{r}{h\sigma_{cd}} + \frac{1}{16}\left(\frac{r}{h\sigma_{cd}}\right)^3\right]
\end{equation}
\noindent in the range $h\sigma_{c} < r \leq 2 h\sigma_{cd}$ and $0$ for $r > 2 h\sigma_{cd}$. We perform both 
one-component simulations at $\rho_d\sigma_d^3h^3=0.27$. We recall that $\rho_d$ is the depletant number density of a 
reservoir at fixed temperature $T$, volume $\mathcal{V}$ and activity $z_d$.

In the full mixture case, the system is in thermal equilibrium with the depletant reservoir, and thus the two share the same 
activity $z_d$. However, the presence of the colloids makes it so that the resulting depletant density of the system is 
not $\rho_d$ but $\rho_d^r = \alpha \rho_d$, where the factor $\alpha=\alpha(q, \rho_c)$ depends solely on the size ratio 
$q$ and the colloid density $\rho_c$. We compute $\alpha$ by means of the free-volume theory~\cite{lekkerkerker_epl}, 
obtaining the value $\alpha = 0.93$. We confirm this value by explicitly computing the excess chemical potential 
$\mu_{\rm ex}$ of the full mixture with the Widom insertion method, since $\alpha = \exp(-\mu_{\rm ex} / k_BT)$. Therefore, 
we simulate a system composed by $N_c=100$ colloids and $N_d=250000$ depletants with size ratio $q=0.1$.
The colloid number density is $\rho\sigma_{c}^3h^3 = 0.1$ and the 
depletant number density is $\rho_d^r\sigma_{d}^3h^3 = \alpha \rho_d\sigma_{d}^3h^3 = 0.25$.

\begin{figure}[ht!]
\includegraphics[width=3.2in]{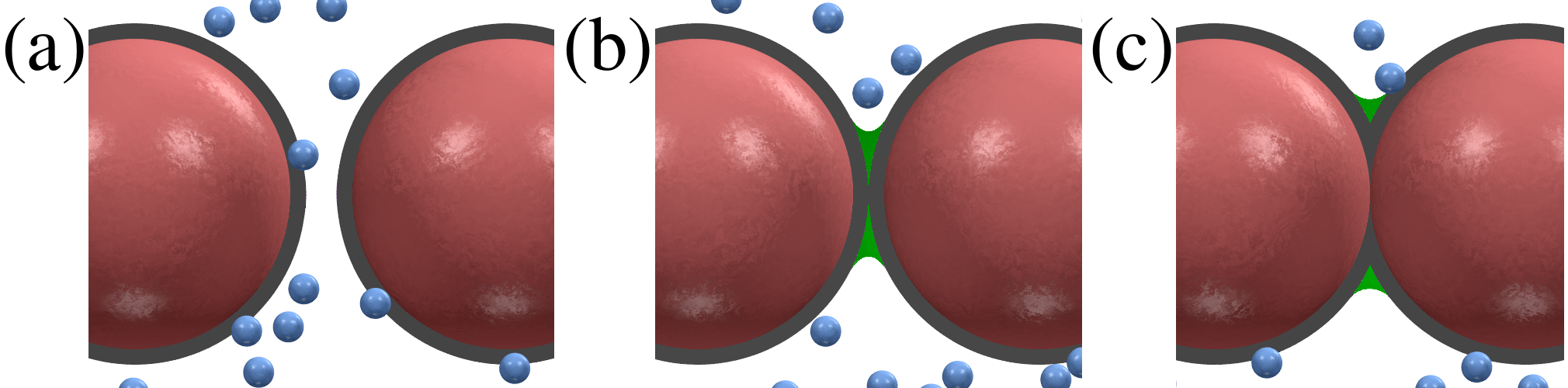}\\
\vspace{0.2cm}
\includegraphics[width=3.2in]{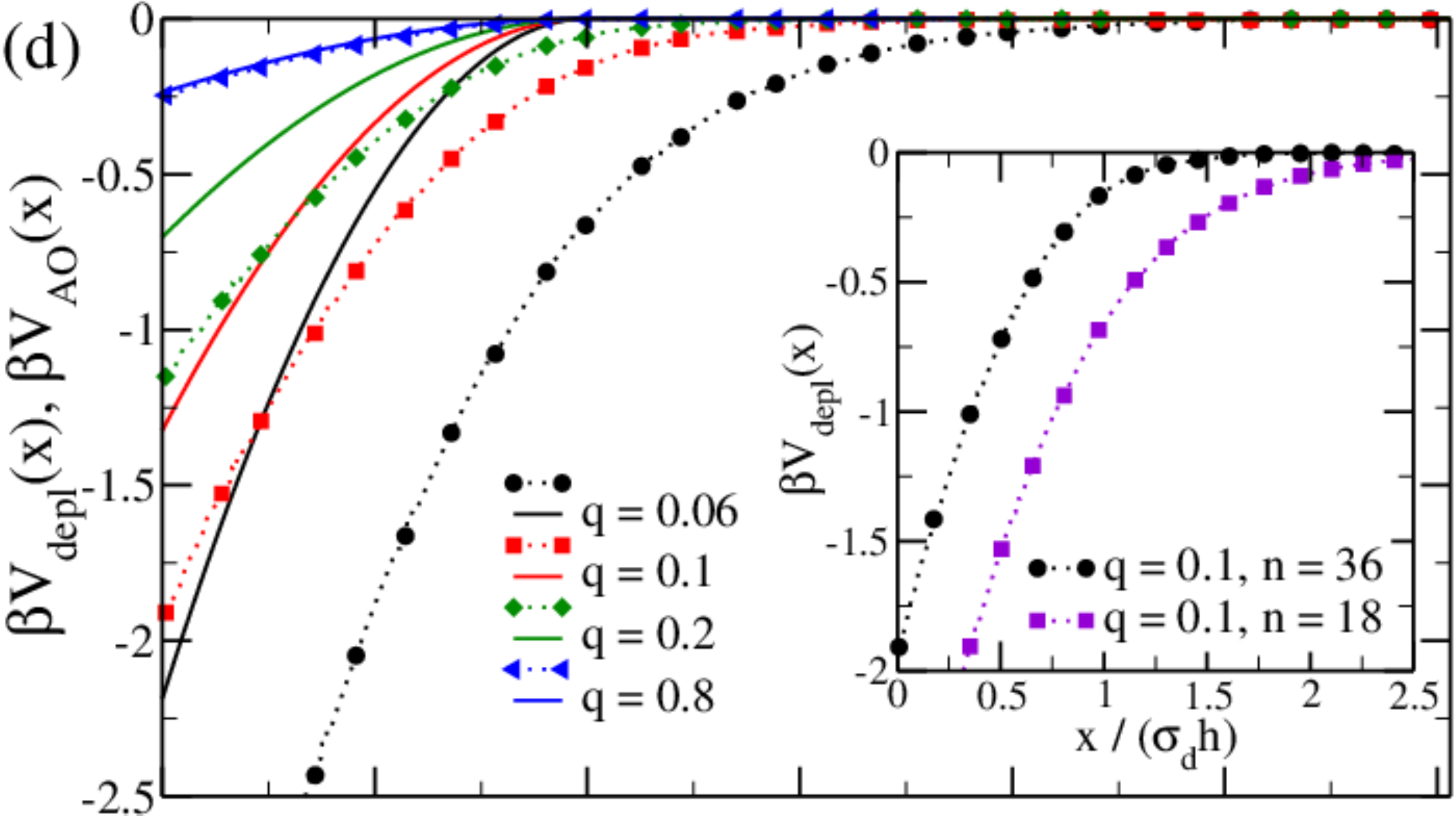}\\
\includegraphics[width=3.2in]{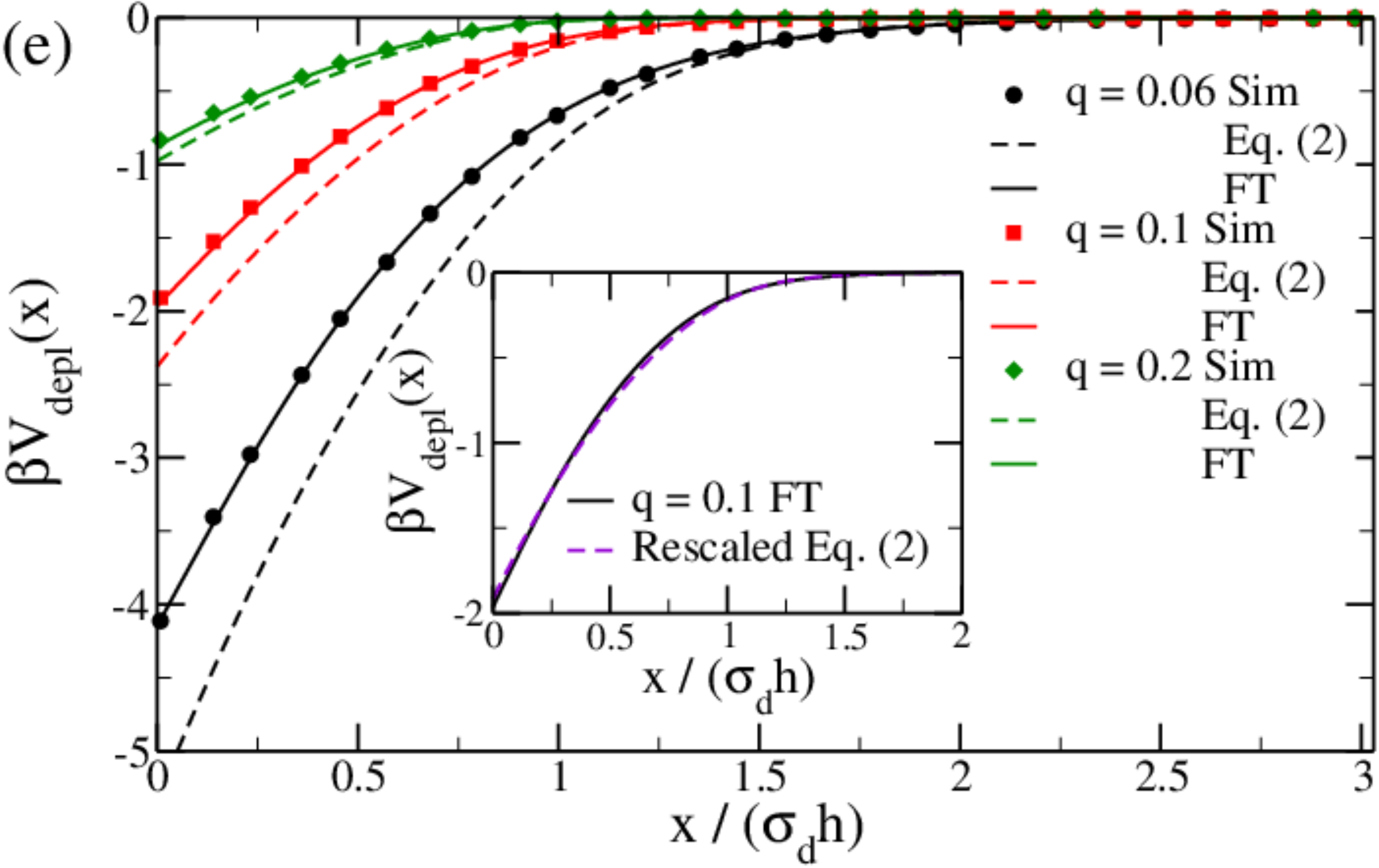}
\caption{(a)-(c): Cartoon of the system. (a) When two colloids (in red) are far apart, depletants (in blue) do not induce 
any interaction. (b) In the case of hard interactions, if the depletion layers (in black) do not overlap, the resulting
effective interaction is null. In the case of soft interactions there is still a depletion area (in green) inducing a non-zero effective attraction. (c) Depletion is enhanced even when the two colloids are very close.
(d): Numerical data for $V_{\rm depl}$ between two large colloids interacting \textit{via} $V_{36}$ both between themselves and with the depletants, as a function of the surface-to-surface colloid distance $x$ and for various values of the size ratio $q$ (dashed lines with points). Full lines show the AO interaction.
Inset: Numerical data for $q=0.1$ at two different values of $n$.
(e): $q$-dependence of $V_{\rm depl}$ for $n=36$ calculated with Eq.~\ref{eq:veff} (full lines), Eq.~\ref{eq:analytic} 
(dashed lines) and numerical data (points). Inset: $V_{\rm depl}$ from Eq.~\ref{eq:analytic} rescaled onto the exact one of Eq.~\ref{eq:veff} using the ratio between contact energies.}
\label{fig:cartoon_n36}
\end{figure}

\section{Results}

\subsection{Effective potentials from soft depletion}
Fig.~\ref{fig:cartoon_n36}(d) reports $V_{\rm depl}$ between two soft colloids for several values of $q$ and fixed depletant density $\rho_d\sigma^3_d h^3=0.158$.
The two colloids interact through $V_{n}$, both between themselves and with the depletants. The latter behave as ideal among themselves. 
As clearly shown, particularly for small $q$, the potential is much more attractive than its AO counterpart. Not only the contact energy decreases, e.g. by more than 30\% in the case $q=0.1$, but, most importantly, it develops a long-distance contribution which grows dramatically upon decreasing $q$. Indeed, while for an AO mixture the range of the interaction is exactly $\sigma_{d}=q\sigma_c$, the corresponding soft version has as exponential tail with a range roughly twice as large in the low-$q$ limit. The AO behaviour is recovered only at large $q$~\citep{zausch2009}. Finally, decreasing $n$ leads to an even larger discrepancy with AO (inset). 
%The corresponding $V_{tot}$ are reported in the SM.

To provide a physical interpretation of these findings, we refer to the cartoons shown in Fig.~\ref{fig:cartoon_n36}(a)-(c). Within the standard AO picture, small particles are excluded from the large particles volume plus the volume of a corona of size $\sigma_d/2$ around them. When the surface-to-surface distance $x$ between large particles is smaller than $\sigma_d$ (and up to contact), the two coronas overlap, giving rise to a larger available volume for the small particles.
In the case of soft interactions particles can $(i)$ partially interpenetrate paying a small energy penalty and $(ii)$ feel a residual repulsion from a colloid even when the latter is relatively distant, due to the potential tail.
It is precisely this tail that causes the effect of increasing the effective attraction among larger soft spheres. Indeed, when large particles are found at $x=h\sigma_d$  
as in Fig.~\ref{fig:cartoon_n36}(b), the depletants that are in the region between the two colloids feel roughly twice the long-tail soft repulsion than those closer to one colloid only. 
As a result, depletants prefer to stay outside the region between the two colloids, thus producing  a residual imbalance in 
the osmotic pressure that generates an attractive force even when colloids are located at a relative distance $x > h\sigma_d$.
The range and strength of the tail are controlled by the functional form of the potential, 
making this effect more pronounced for softer particles, while naturally recovering AO for hard particles, i.e. for $n\rightarrow \infty$ (see Appendix). Since this enhanced depletion depends entirely on the behaviour of the soft tail of the interaction potential, it is expected to be generic for any soft system.
In addition, this effect is important only for small $q$-values, because the range of the tail must effectively compete with the range of the effective potential.  Thus, for larger $q$ the potential is typically negligible even at a surface-to-surface distance $\sigma_d$, so that 
one still recovers the AO behaviour even for small $n$. This explains the findings of Zausch {\it et al}~\cite{zausch2009}, who showed the equivalence between a soft and the AO mixture for $q=0.8$.
For small $q$ values, on the other hand, our results show that a mapping between AO and soft depletion is not possible. 
We believe that this large effect of softness on the interaction potential between colloids could be experimentally observed through different experimental approaches. For instance, a direct way to obtain the potential mean-force can be achieved through the use of holographic microscopy~\cite{fung_holography}, which allows 
 to compute the histogram of distances (or, equivalently, the radial
distribution function), of a diluted colloidal system in the presence of soft depletants. Also, a possible way to establish the effective depletion potential is by using confocal microscopy even in more dense colloidal systems, to compare the measured radial distribution function with theoretical/numerical predictions~\cite{mohanty_pnipam}.

We now show that Eq.~\ref{eq:veff} quantitatively describes the numerical results. The comparison
between the theoretical and simulation data is reported in Fig.~\ref{fig:cartoon_n36}(e), showing that the two sets fall on top of each other. In addition, we also plot the predictions obtained by the asymptotic formula in Eq.~\ref{eq:analytic}, which applies to the case of inverse power-law pair potentials.  Despite a systematic overestimation of the contact energy, the description of the data is qualitatively correct. Moreover, scaling the predictions to the contact energy of the exact results provides a very good agreement (inset of Fig.~\ref{fig:cartoon_n36}(e)), thus proving that soft depletion in this case has a $\sim \exp(-n\,x/\sigma_{cd})$ tail for $x/\sigma_d >1$ which is essentially controlled by $n$ (and to a smaller extent by $q$). Therefore, this contribution is always present for finite $n$, and one can never recover the AO result, e.g. by a simple rescaling of $\rho_d$. 

\begin{figure}[ht!]
\includegraphics[width=3.2in]{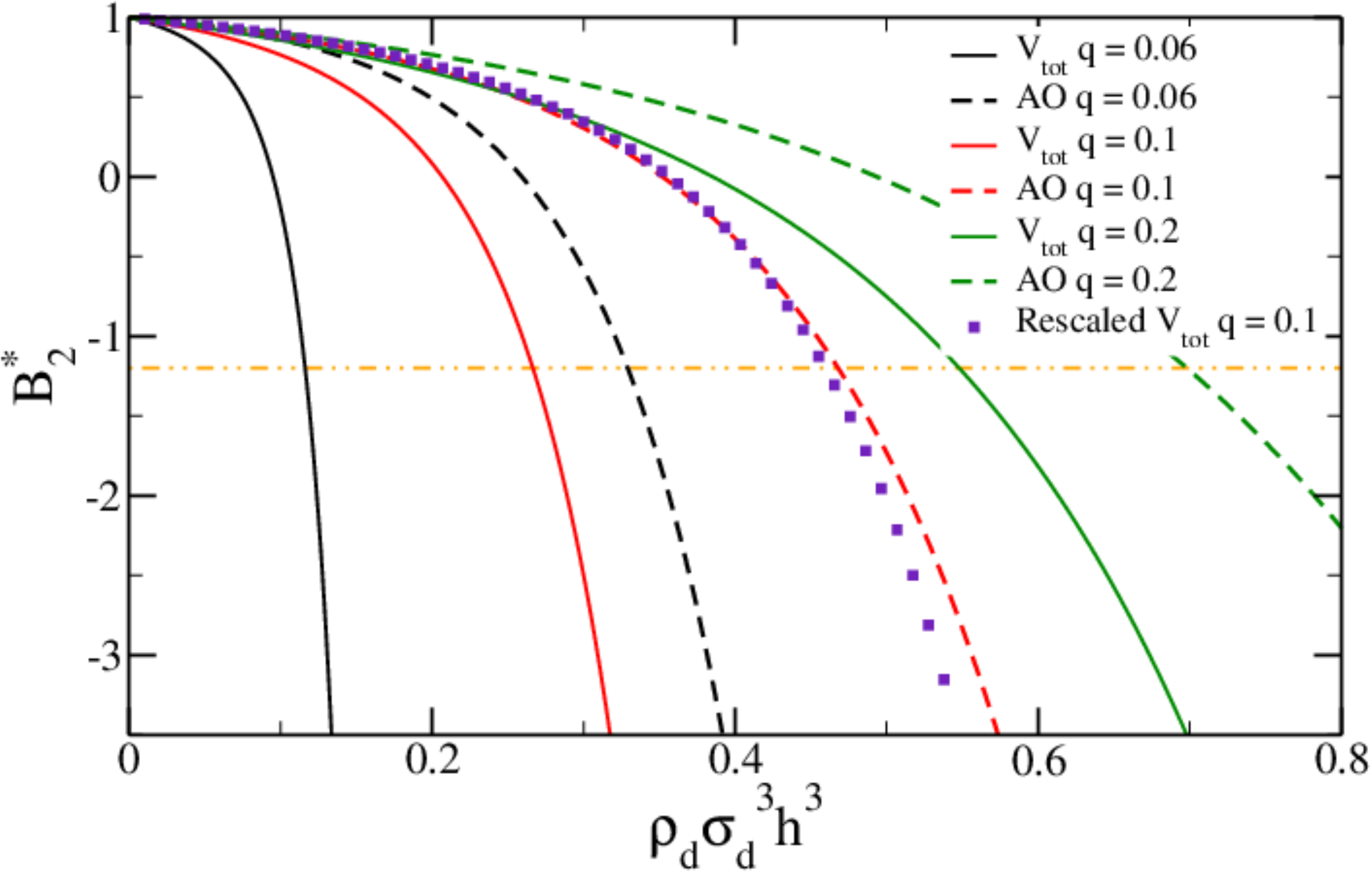}
\caption{Normalized second virial coefficient $B_2^*$ as a function of $\rho_d\sigma^3_d h^3$ for $V_{\rm tot}=V_{36}+V_{\rm depl}$ (full lines) and for the AO case (dashed lines). The dashed-dotted horizontal line indicates where colloids should phase separate according to the extended law of corresponding states. 
Filled squares are data for $V_{\rm tot}$ at $q=0.1$ as a function of the depletant density, rescaled by a factor $\approx 1.72$ that best fits the AO case, to highlight the different $\rho_d$ dependence between the two. For $n=36$, $h=1.01818$\protect\cite{inverse_B2}.
}
\label{fig:B2}
\end{figure}

\subsection{Consequences on colloidal phase behaviour}
As we mentioned before, potentials of different shape behave in an identical way when compared using $B_2^*$ as a control parameter~\cite{noro_frenkel}. Indeed, it was shown that such systems exhibit a gas-liquid phase separation at $B_2^*\sim -1.2$~\cite{noro_frenkel,Mil03a,buzzaccaro_B2,LuCiulla,LuCiulla2,gibaud_b2,secchi14}.

We calculate $B_2^*$ for the total potential among colloids, $V_{\rm tot}=V_{n}+V_{\rm depl}$, for different $q$ values as a function of $\rho_d$. 
The results, reported in Fig.~\ref{fig:B2}, clearly show the dramatic consequences of softness, which leads to a significant decrease of the critical $\rho_d$ with respect
to the AO case.
For small $q$, this decrease is of the order of a factor 2. Only for large $n$ the AO behaviour is recovered (see Appendix). However, for small and intermediate $n$, a simple rescaling of the density or of the size ratio is not sufficient to reproduce the AO behaviour in the whole $\rho_d$ range. Indeed, leaving $\rho_d$ as a free 
parameter is not sufficient to rescale the soft and AO curves on top of each other, as shown by the purple squares in Fig.~\ref{fig:B2}. A perfect curve 
collapse is unattainable even for different soft potential, i.e. inverse power law interactions with distinct $n$ exhibit different $B_2^*$ $\rho_d$-dependences owning
to the peculiar functional forms of the resulting $V_{\rm tot}$.

\begin{figure}[ht!]
\begin{tabular}{lr}
\includegraphics[width=1.9in]{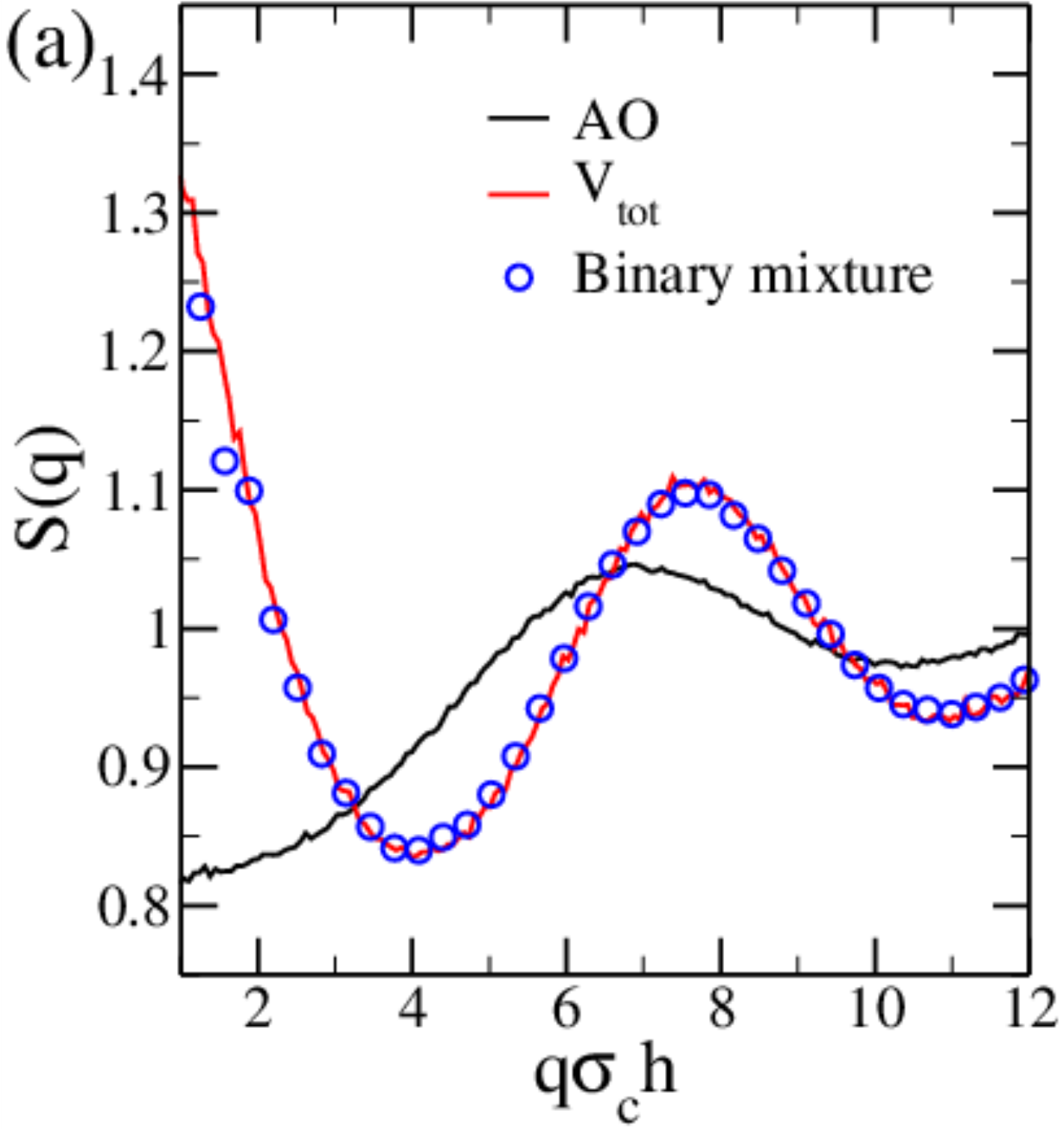} & \includegraphics[width=1.2in]{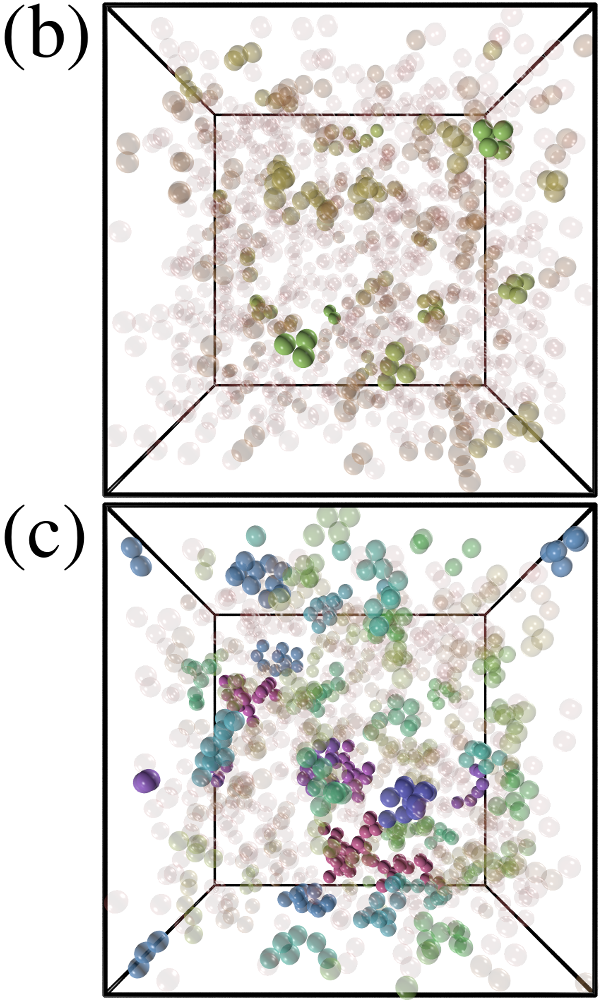}
\end{tabular}
\caption{(a) Static structure factor $S(q)$ computed for a system of colloids at density $\rho_c\sigma_c^3 h^3 = 0.1$, interacting via AO (black) and $V_{\rm tot}$ (red) with $n=36$ at $\rho_d\sigma_d^3 h^3=0.27$ ($B_2^* \sim -1$). The corresponding full mixture (blue) is simulated at a depletant density $\rho_d^r\sigma_d^3 h^3 = 0.25$ in order to account for the diminished free-volume fraction available to the depletants~\cite{lekkerkerker_epl}. Panels (b) and (c) show snapshots of the AO (b) and soft (c) one-component systems. Particles are coloured according to the size of the cluster they are part of. Monomers and small clusters are transparent to make larger clusters stand out.}
\label{fig:sq}
\end{figure}

The knowledge of the $\rho_d$-dependence of $B_2^*$ becomes crucial in order to simulate the full mixture, a practice that is becoming more and more common thanks to the increasing usage of GPU computing~\cite{owens2008gpu,nickolls2010gpu,colberg_gpu,rovigatti_gpu}. While the thermodynamics and structure  of the depletants can be investigated \textit{via} MC simulations by employing non-local update algorithms~\cite{cluster_algorithm,wilding_asymmetric}, brownian dynamics simulations also allow to assess the dynamics of depletants and colloids.
To this aim, we consider an effective one-component colloidal system interacting with $V_{\rm tot}$ and compare it with the corresponding soft binary mixture.
We find perfect agreement between the two, as shown in Fig.~\ref{fig:sq}(a), where the two colloidal structure factors fall on top of each other. Incidentally, this comparison also shows that many-body forces are negligible and the two body effective potential given by Eq.~(\ref{eq:veff}) suffices for the study of the equilibrium properties of the suspension at small $q$. On the other hand, for the AO potential at the same $\rho_d\sigma_d^3 h^3$ the system lives at a much higher $B^*_2$, as also shown by the representative snapshots in Figs.~\ref{fig:sq}(b)-(c). 

\subsection{Generality of the results: the case of hertzian particles}
\begin{figure}[!ht]
\includegraphics[width=3.2in]{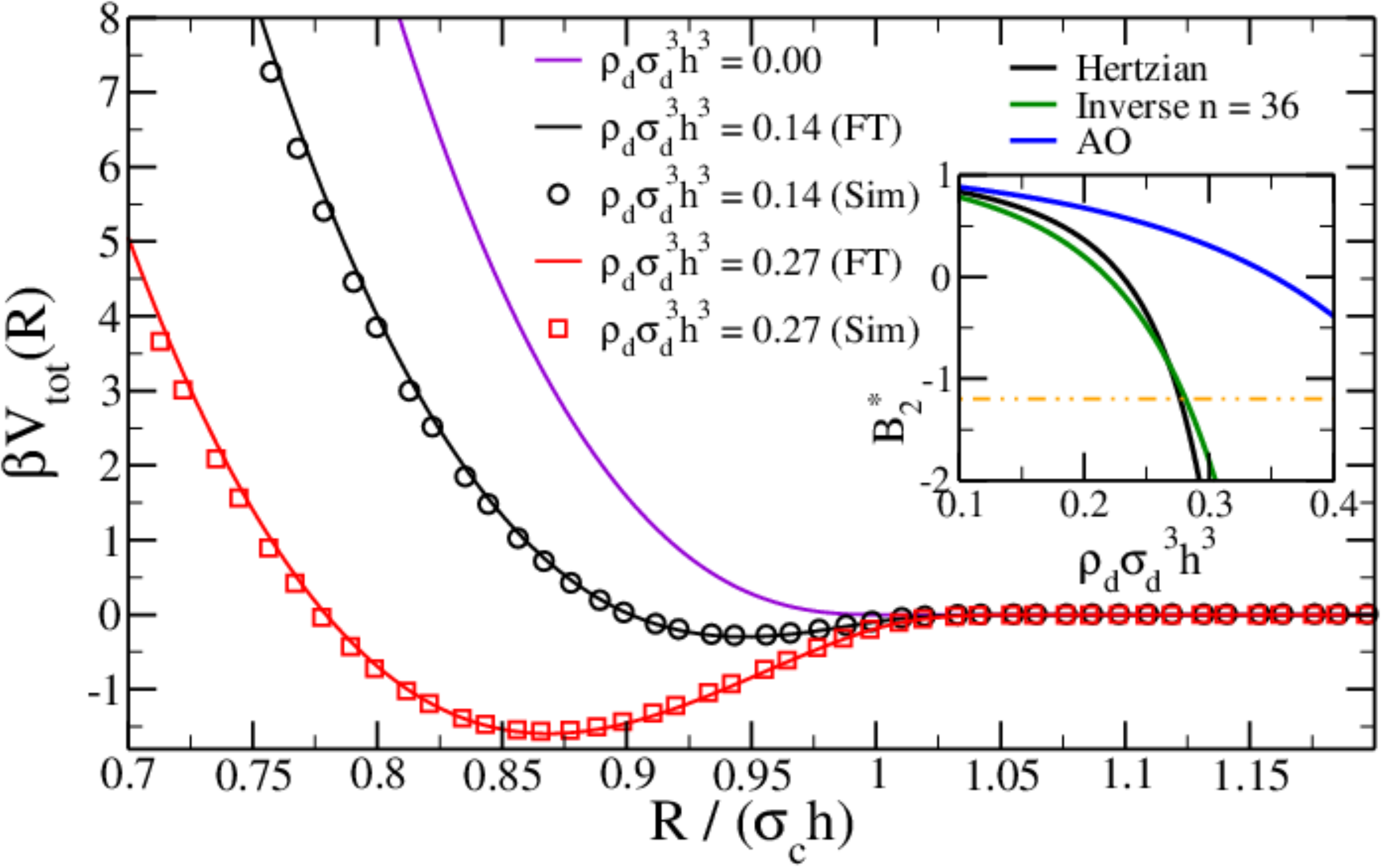}
\caption{$V_{\rm tot}$ for two hertzian particles at a center-to-center distance $R$ immersed 
in a sea of ideal depletants with $q=0.1$ at different depletant densities. Colloid-colloid and colloid-depletant interactions are hertzian with strength $\lambda=500$ $k_BT$. Lines are theoretical results from Eq.\protect\ref{eq:veff} and points are simulation data. Inset: second virial coefficient $B_2^*$ as a function of depletant density for the hertzian mixture and comparison with the AO (blue line) and $n=36$ (green line)  cases at $q=0.1$. For the studied $V_{HZ}$ $h=0.9272$. The horizontal line marks the phase separation threshold.
}
\label{fig:microgel}
\end{figure}

To show that the above results hold for any type of soft particles, we also consider particles interacting with a Hertzian potential $V_{HZ}$. Such a potential well describes the fluid structure of microgel suspensions~\cite{mohanty_pnipam}. We use $V_{HZ}$ for both colloid-colloid and colloid-depletant interactions, while depletants are ideal among themselves. 
In Fig.~\ref{fig:microgel} we show $V_{\rm tot}$, calculated from simulations and using Eq.~\ref{eq:veff} for $V_{\rm depl}$, at $q=0.1$ and for various values of $\rho_d$. The agreement is quantitative. 
As in the case of $V_n$, we find that the range of the total resulting potential, defined as 
 the width of the attractive part of the potential, is larger than $q$ and grows considerably as $\rho_d$ increases, with the attractive minimum moving to smaller and smaller distances as particles start to interpenetrate. The inset of Fig.~\ref{fig:microgel} also
 shows $B_2^*$ calculated for $V_{HZ}$ in comparison to that for $V_{36}$ and for the AO case at the same value of $q$, confirming 
the very different nature of soft depletion with respect to the AO model.

\subsection{Effect of soft repulsion among depletants}
At this point, it appears important to address the role of depletant-depletant interactions on the results presented in the previous sections.
In particular, it would be desirable to get some insights as to whether soft repulsions, naturally arising between the depletant themselves, can act against the enhancement of attraction induced by softness with respect to the AO case, and thus reduce the practical importance of our results for real systems.
The most used depletants are polymers, which can be modelled in a coarse-grained treatment as weakly repulsive gaussian particles~\cite{louis_prl}. Therefore, one would expect that their mutual interactions do not significantly modify our findings with respect to the ideal case.
On the other hand, it is well known that, in the case of hard sphere depletants, the effective potential develops oscillations, modulated by the depletant size, 
causing on top of the attraction a repulsion among colloids at certain distances\cite{Likos}. One may then be inclined to think that a repulsion among depletants may weaken the total effective attraction among colloids. To investigate whether this is the case, we have performed additional simulations to calculate the effective potential in the case of two soft colloids, modelled with $V_{36}$, immersed in a sea of (i) polymers, interacting with a gaussian potential following Bolhuis and Louis~\cite{bolhuis_macromol}, and (ii) soft particles, also interacting with $V_{36}$. In both cases, the cross interactions between colloids and polymers are also modelled with $V_{36}$ and the size ratio has been fixed to $q=0.1$.

\begin{figure}[!ht]
\includegraphics[width=3.2in]{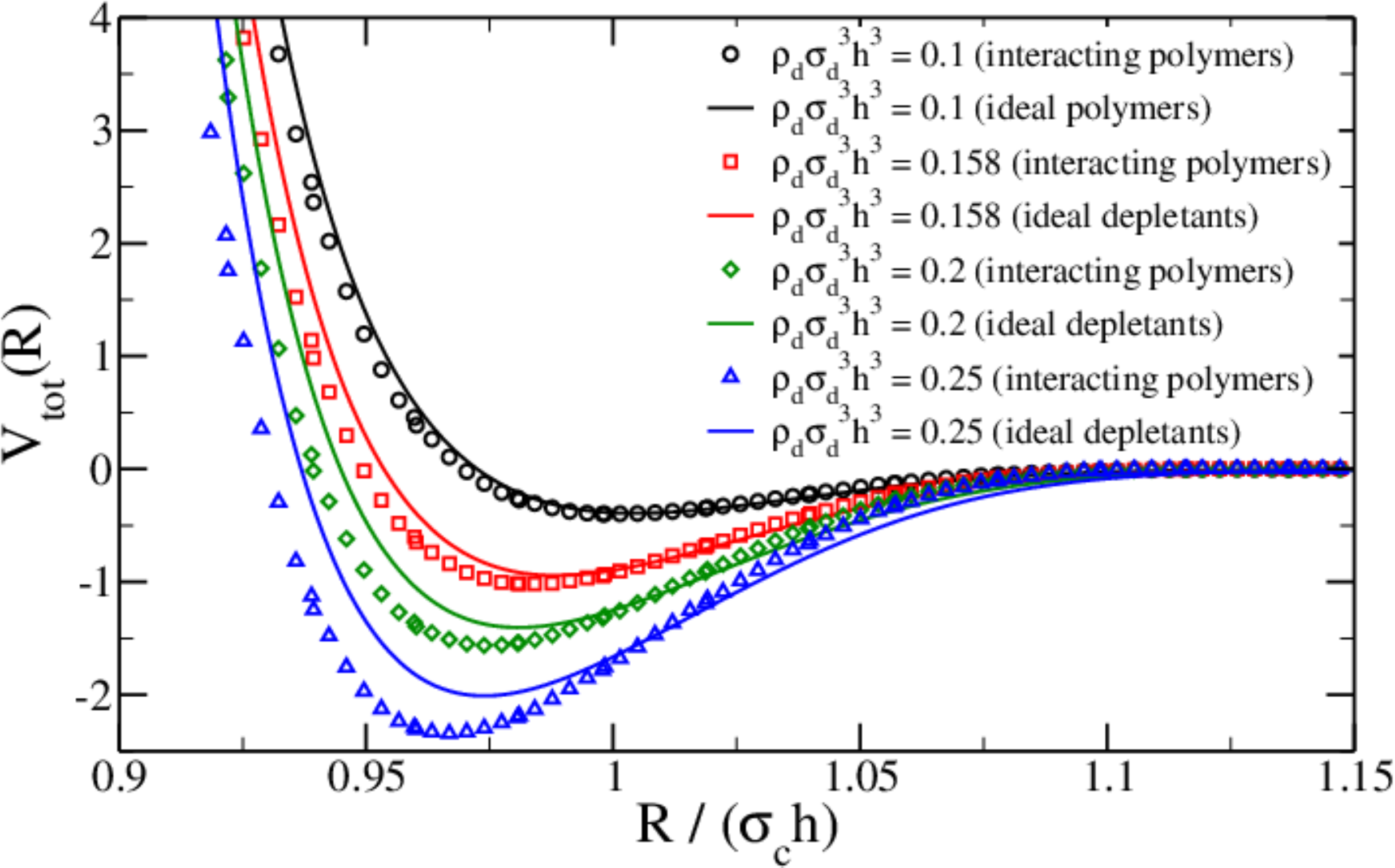}
\caption{$V_{\mathrm{tot}}$ in the presence of ideal depletants (solid lines) and interacting polymers (symbols) for different values of depletant density and for $q=0.1$.
The potential among polymers is fixed following Ref.\protect\cite{bolhuis_macromol}, amounting to a Gaussian potential with amplitude $2k_B T$ and variance $0.6\sigma_p$.}
\label{fig:Intpol}
\end{figure}

\begin{figure}[!ht]
\includegraphics[width=3.2in]{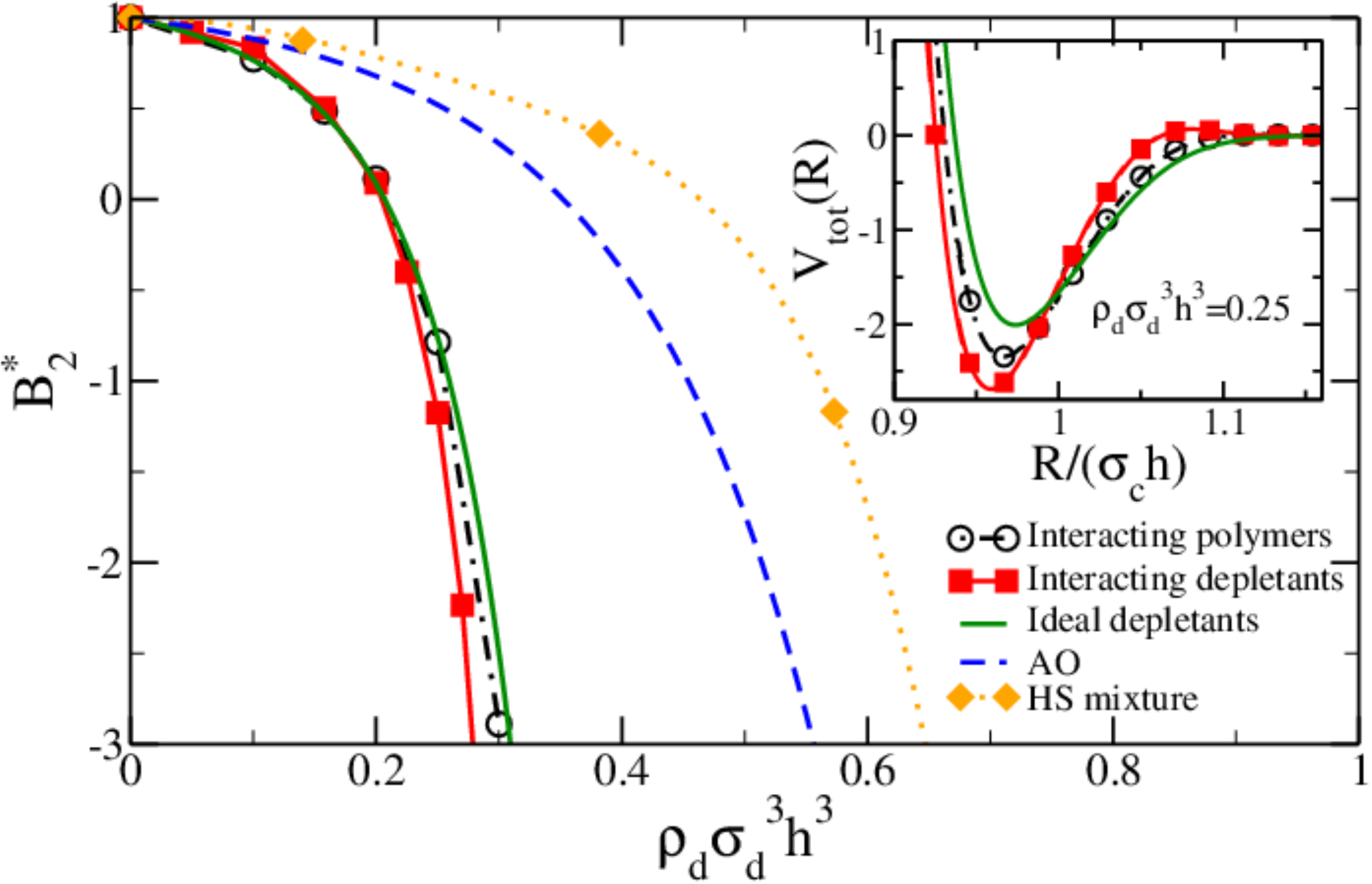}
\caption{$B_2^*$ as a function of depletant density in the presence of ideal depletants (solid lines), interacting polymers (open circles) and interacting soft depletants (full squares)  compared with the AO model (dashed line) and the HS mixture (full diamonds) for $q=0.1$. The lines connecting symbols are guides for the eye. Regarding the HS mixture data, the data point at $\rho_d\sigma_d^3 h^3=0.14$ is obtained from the effective potential evaluated in Ref.\protect\cite{Nico_SM}, while the rest of the points are obtained by data in Ref.\protect\cite{roth_pre}. Inset: $V_{\mathrm{tot}}$ for the ideal, polymer-like and soft depletant cases at fixed $\rho_d\sigma_d^3 h^3=0.25$.}
\label{fig:SmallInt-B2}
\end{figure}

Fig.~\ref{fig:Intpol} shows $V_{\mathrm{tot}}$ among colloids in solution with interacting polymers, compared to the case where depletants are ideal for different depletant density. As expected, the deviations from ideal behaviour are more evident as $\rho_d$ increases. More specifically, the attraction reduces at large distances, but the effect appears to be compensated by a deeper minimum in the potential, which moves at shorter distances with respect to the ideal case. 
Such an effect is generated by the softness of the colloids, which allows for a deeper interpenetration.
To quantify the effect of the polymer-polymer interaction on the colloidal phase behaviour, we numerically compute the normalised second virial coefficient for the investigated state points. This is shown in Fig.~\ref{fig:SmallInt-B2}. We can thus conclude that, while deviations in the potential are observed also for not so high values of the depletant density, $B_2^*$ does not change much due to the compensation between increased attraction at contact and increased repulsion at distances $R/(\sigma_c h) \sim 1+q$.

In the case of depletants that are not as soft as polymers, more significant deviations from the ideal case are observed. This is shown in the inset of Fig.~\ref{fig:SmallInt-B2}.
Here, we compare $V_{\mathrm{tot}}$ for the ideal case both with Gaussian polymers and with soft depletants interacting with $V_{36}$. The effects already noticed for polymers are still present, being actually amplified by a stronger depletant-depletant repulsion. Thus, a (repulsive) peak develops at large distances, while the minimum at contact becomes larger and moves to smaller distances. The resulting $B_2^*$, also shown in  Fig.~\ref{fig:SmallInt-B2}, tells us that the increased attraction is dominant, further lowering the second virial coefficient with respect to the ideal case. Thus, even in the presence of depletant interactions in the form of soft repulsion, we find that the AO model remains very far from what observed in soft mixtures and depletion is always enhanced by softness. 
However, one may argue that a system composed of small and large soft spheres (with all interactions modelled as $V_{36}$) should be naturally compared to a mixture of hard spheres, rather than to the AO model. To this aim we complement Fig.~\ref{fig:SmallInt-B2} with calculations of $B_2^*$
for a $q=0.1$ HS mixture. We find that in this case the second virial coefficient becomes negative at a much larger depletant density not only with respect to the 
 the soft mixture case, but also with respect to the AO mixture~\cite{roth_pre,roth_hs_mixture}\footnote{We note that the use of the simple formula for the effective depletion potential in HS mixtures proposed by Gotzelmann {\it et al} \protect\cite{Gotzelmann,dijkstra_1999} gives results for the second virial coefficient in between the AO mixture and the ones for data from Roth and coworkers \protect\cite{roth_pre}.}. Therefore,
we conclude that the enhanced attraction seen in the systems under study comes unequivocally from the soft nature of the colloid-polymer (and colloid-colloid) interaction.

\section{Conclusions}
In this work we have established that, for small depletant-to-colloid size ratio, soft depletion cannot be described by a suitable rescaling of the AO potential. Indeed, the tail of soft interactions enhances, both in depth and in range,  the effective colloid-colloid attraction induced by soft depletants of small sizes. We have also provided a simple theoretical expression which allows us to evaluate the effective potential among colloids and ideal depletants of any mutual interaction.  These findings hold for different models of repulsive soft particles, having consequences on the phase behaviour of colloidal systems, which is usually established by means of effective one-component potentials. When these take into account the softness of the particles, phase separation requires a much lower density of depletants with respect to the AO case. This makes it unfeasible to neglect softness when mapping the full mixture of colloids and co-solutes onto the effective one-component system. Corrections arising for non-ideal depletants do not qualitatively alter our findings, confirming the enhancement of attraction among colloids also when depletants interact among themselves with soft repulsion. 
 %will not significantly alter our findings, since colloidal phase separation at small $q$ occurs below the overlap concentration~\cite{bolhuis_prl}. Indeed, preliminary results show that adding repulsion between the depletants for the inverse potential interaction case does not sensibly alter the resulting $V_{\rm eff}$ for small $q$-values up to $\rho_d$ values where the system is expected to phase separate. However, the effect of non-ideality in the case of PNIPAM microgels seem to be larger, but smaller in magnitude with respect to the overall effect of softness. More work on this topic is underway.
Finally, we stress that our representation of ``soft'' spheres interacting through a $n=36$ inverse power-law potential is often used as a ``hard'' approximation \cite{puertas_zaccarelli,puertas_n36,melrose_n36} and, most importantly, 
that real colloidal particles are typically at least as soft. In addition, depletion forces are not only relevant for colloidal science, but also play an important role in biological and cellular organisation, typically involving soft particles~\cite{marenduzzo,sarkar_bio_depletion}.
Thus, our results have a broad generality and call for the need to go beyond the AO model when dealing with more realistic depletion effects due to repulsive forces.

\emph{Acknowledgments} LR acknowledges support from the Austrian Research Fund (FWF) through his Lise-Meitner Fellowship M 1650-N27. NG, EZ acknowledge support from MIUR ("Futuro in Ricerca" ANISOFT/RBFR125H0M). We thank I. Coluzza, C. Likos and P. Schurtenberger for helpful discussions.

\bibliography{main_text}

\section{Appendix}

%\subsection{Simulation techniques}

\subsection{Analytical approximation}
Here we derive the approximate expression, given by Eq.~\ref{eq:analytic}, of the effective 
interaction for an inverse power-law potential. We obtain an analytical form when $n$ is large,
$q$ is small and $nq \gtrsim 1$. 
By use of the convolution theorem, the Fourier transform of the effective interaction reads
\begin{equation}
-\beta\,V_{\rm depl}(k) = \rho_d\,\left [ \int\,d{\bf r} \left ( e^{-\beta\,v_{cd}(r)}-1\right )\,
e^{i{\bf k}\cdot{\bf r}} 
\right ]^2 
\end{equation}
We first note that for $|r-\sigma_{cd}| \ll \sigma_{cd}$ (i.e. $q \ll 1$)
\begin{equation}
v_{cd}(r) = \left (\frac{1}{1+\frac{r-\sigma_{cd}}
{\sigma_{cd}}}\right )^n \sim e^{-n\frac{r-\sigma_{cd}}{\sigma_{cd}}}
\end{equation}
In the regime of interest, the Mayer function can be roughly approximated as
\begin{equation}
e^{-\beta\,v_{cd}(r)}-1 \sim 
\begin{cases}
-1 & {\rm for} \quad r < \sigma_{cd}
\cr
-\beta v_{cd}(r) & {\rm for} \quad r > \sigma_{cd}
\end{cases}
\end{equation}
In this limit, which becomes exact for $n\to\infty$,
the Fourier transforms can be evaluated analytically and the effective potential is given by
Eq. (\ref{eq:analytic}) with 
\begin{eqnarray}
Q_1 &=& A + B\,x + C\,x^2 + \frac{\sigma_{cd}}{3}\,x^3
+\frac{x^4}{24} \\
Q_2 &=& A + B^\prime\,x+ C^\prime \,x^2 +\frac{x^3}{6\alpha}
\end{eqnarray}
with $x=R-2\sigma_{cd}$  and $\alpha= \frac{n}{\sigma_{cd}}=\frac{2nq}{1+q}$.
The explicit expressions of the parameters are:
\begin{eqnarray}
A &=& \frac{3}{\alpha^2}\sigma_{cd}^2 + \frac{4}{\alpha^3}\sigma_{cd}- \frac{5}{\alpha^4}
 \\
B &=& \frac{4}{\alpha^3} - \frac{2}{\alpha}\sigma_{cd}^2
 \\
C &=& \frac{\sigma_{cd}^2}{2} - \frac{\sigma_{cd}}{\alpha} - \frac{1}{\alpha^2}
 \\
B^\prime &=& \frac{\sigma_{cd}^2}{\alpha} + \frac{4}{\alpha^2}\sigma_{cd} - \frac{1}{\alpha^3}
 \\
C^\prime &=& \frac{\sigma_{cd}}{\alpha} + \frac{1}{2\alpha^2}
\end{eqnarray}
This effective interaction tends to the Asakura-Oosawa form in the limit $n \to\infty$.

\subsection{Inverse-power-law $V_{\rm tot}$ as a function of $q$ and as $n \to \infty$}
\begin{figure}[ht!]
\includegraphics[width=3.2in]{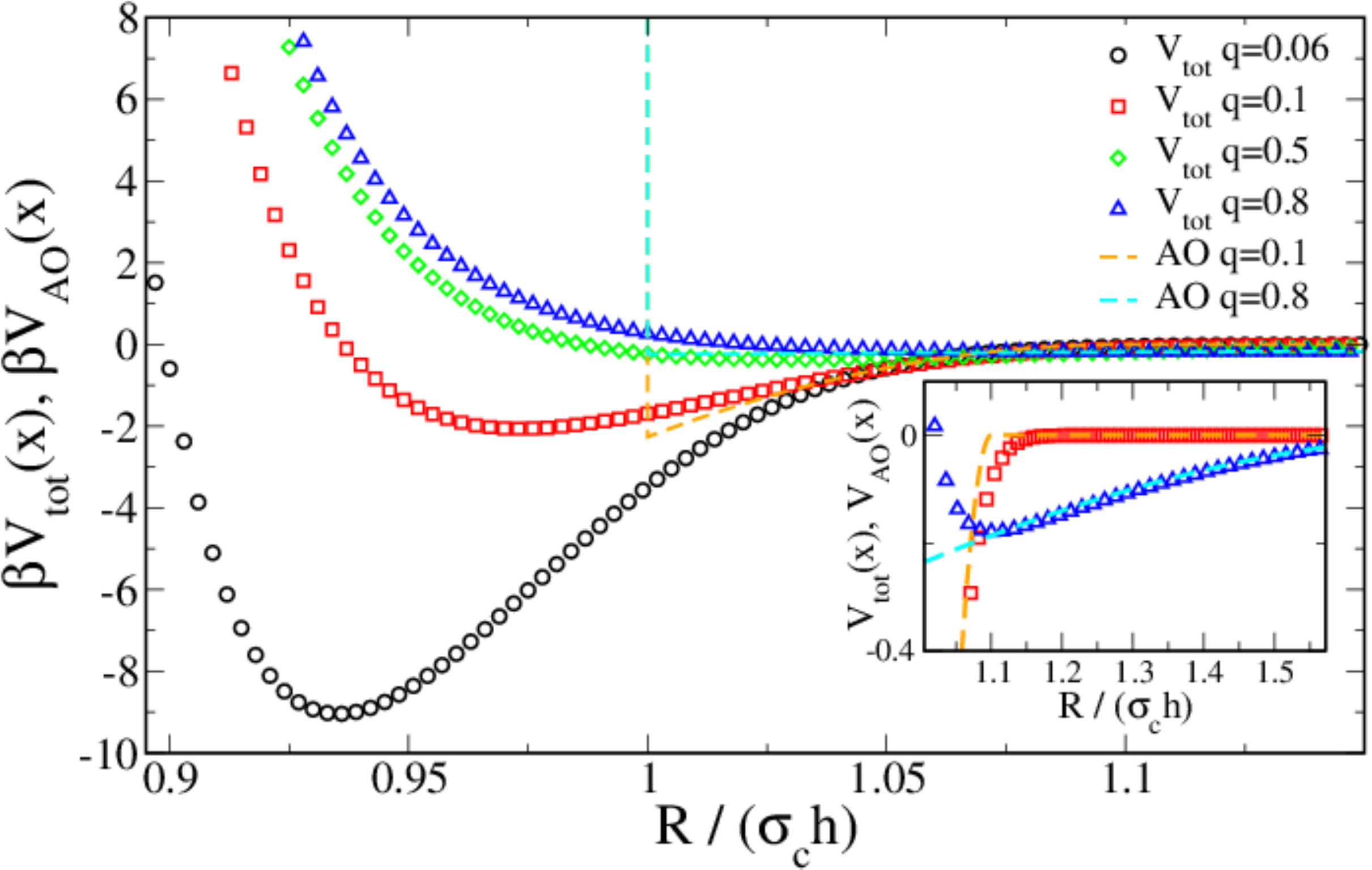}
\includegraphics[width=3.2in]{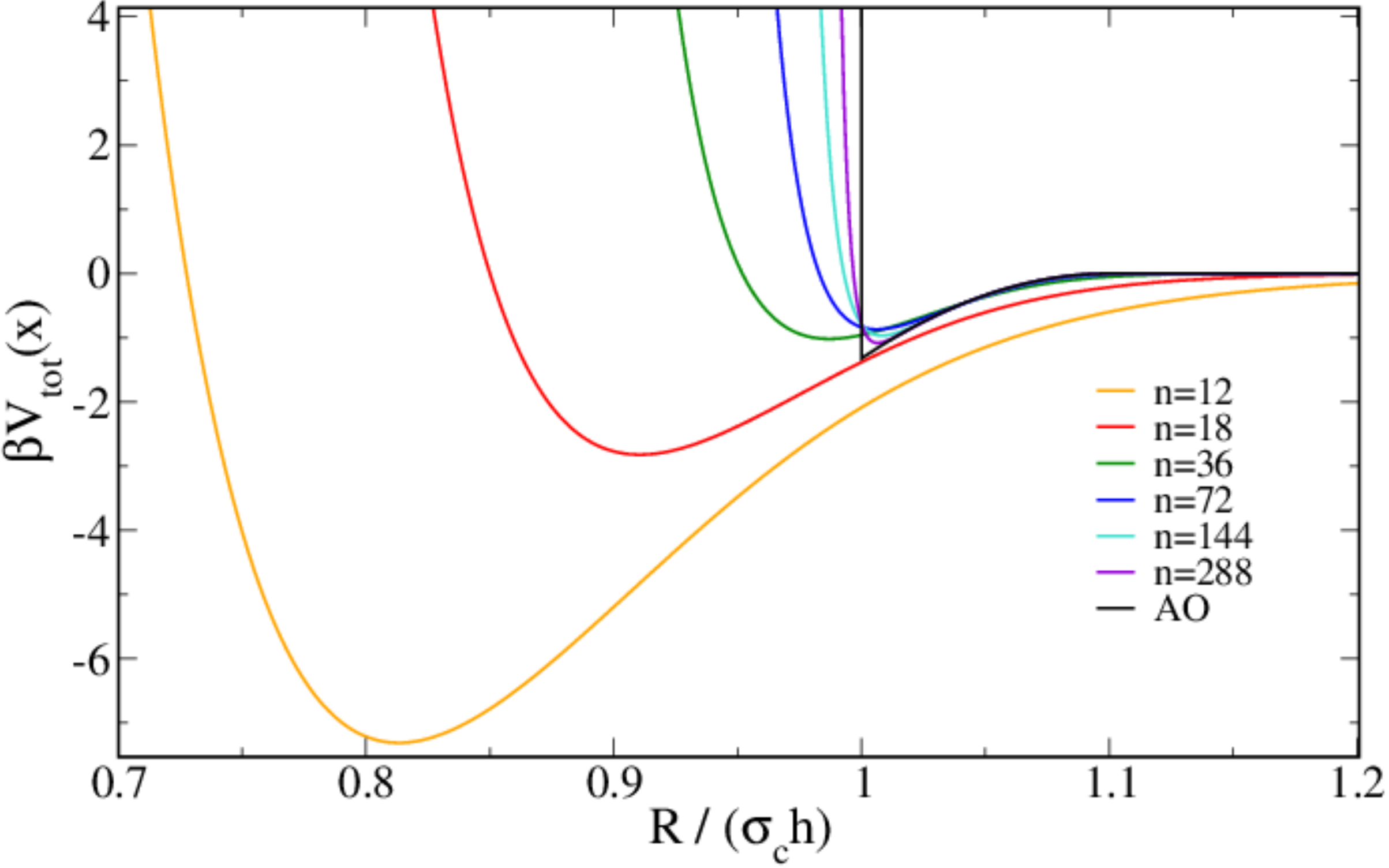}
\caption{$V_{\rm tot} = V_{n} + V_{\rm depl}$ for a mixture of soft colloids and depletants for $\rho_d\sigma^3_d h^3=0.158$ and for (top) different values of $q$ and $n=36$. In the inset a zoom of the data is provided to see that at large $q$, the AO behaviour is recovered; (bottom) $q=0.1$ and different values of $n$. Here, as $n \to \infty$, $V_{\rm tot} \to V_{HS} + V_{AO}$ (black line).  The dependence of $h$ on $n$ is described in Ref.\protect\cite{inverse_B2}. }
\label{fig:Vtot}
\end{figure}

\begin{figure}[ht!]
\includegraphics[width=3.2in]{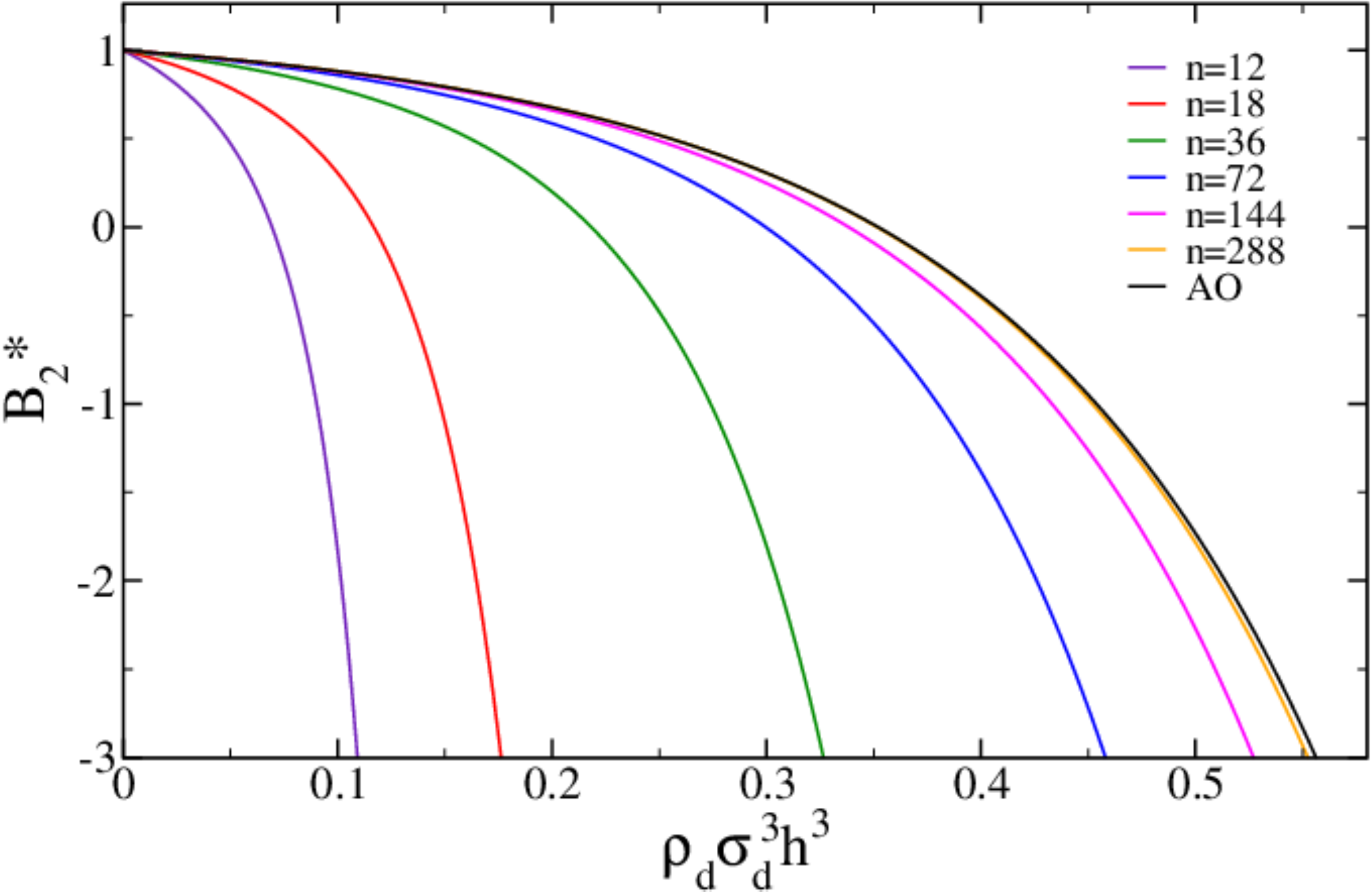}
\caption{$B_2^*$ for different $n$ for $V_{\rm tot}$ (small and large particles behaving as inverse power-law pair potentials) and $V_{AO}$  as a function of depletant density and $q=0.1$. }
\label{fig:n}
\end{figure}

Here we show how the AO limit is recovered as $n\to \infty$.
Fig.~\ref{fig:Vtot} shows the $q$-dependence (top) and $n$-dependence (bottom) of $V_{\rm tot}(r) = V_n(r) + V_{\rm depl}(r)$.
Fig.~\ref{fig:n} reports the $n$-dependence of the normalised second virial coefficient $B_2^*$ as a function of the depletant density $\rho_d$.
The AO limit can be strictly recovered only for $n\rightarrow \infty$. For small $n$, at large $q$, the AO behaviour is also recovered (except for the soft core)
and the range of $V_{tot}$ and $V_{AO}$ become the same, so that the two can be mapped onto each other\cite{zausch2009}. However, in this limit, a one-component picture becomes questionable and many-body forces become important.

\end{document}